\documentclass[pre,showpacs,twocolumn,preprintnumbers,amsmath,amssymb,footnotebib]{revtex4-1}
\usepackage{epsfig}
\usepackage{graphicx}
\usepackage{textcomp}
\usepackage{tabularx}

\usepackage{empheq}
\usepackage[mathscr]{euscript}

\begin{document}
\title{FKPP dynamics mediated by a parent field with a delay.}

\author{Steffanie Stanley}
\affiliation{California Polytechnic State University, San Luis Obispo, NY, 93407}

\author{Oleg Kogan}
\affiliation{California Polytechnic State University, San Luis Obispo, NY, 93407}
\email{okogan@calpoly.edu}

\begin{abstract}
We examine a modification of the Fisher-Kolmogorov-Petrovsky-Piskunov (FKPP) process in which the diffusing substance requires a parent density field for reproduction.  A biological example would be the density of diffusing spores (propagules) and the density of a stationary fungus (parent).  The parent produces propagules at a certain rate, and the propagules turn into the parent substance at another rate.  We model this evolution by the FKPP process with delay, which reflects a finite time typically required for a new parent to mature before it begins to produce propagules.  While the FKPP process with other types of delays have been considered in the past as a pure mathematical construct, in our work a delay in the FKPP model arises in a natural science setting.  The speed of the resulting density fronts is shown to decrease with increasing delay time, and has a non-trivial dependence on the rate of conversion of propagules into the parent substance.  Remarkably, the fronts in this model are always slower than Fisher waves of the classical FKPP model.  The largest speed is half of the classical value, and it is achieved at zero delay and when the two rates are matched.  
\end{abstract}

\maketitle

\section{Introduction}
Many microorganisms reproduce by means of spores.  Spores are produced by a parent organism \cite{Britannica}, and have a probability to produce a new parent organism.  After some delay - known as latent time - a new parent organism will begin to produce new spores.  Fungi are an obvious example of this process \cite{FungalBio1} (see also Chapter 11 of \cite{FungalBio2}), but reproduction by means of spores occurs also among algae and plants 
\cite{Britannica}.  Slime molds also employ spores in their life cycle \cite{SlimeMold}.   The spores are an example of what's called ``propagules'' in ecology \cite{Propagules}.  

The population dynamics of reproducing microorganisms often gives rise to invasion fronts \cite{FKPP1}-\cite{FKPP2}.  The dynamics of invasion fronts of organisms that reproduce by direct division can be described by the Fisher-Kolomogorov-Petrovskiy-Piscunov (FKPP) model \cite{FKPP1}-\cite{FKPP2}.  This well-known model originated in the context of population genetics \cite{FKPP3}-\cite{FKPP4}, but has found applications in fields as diverse as biochemical waves during development to nuclear physics \cite{FKPP2}, \cite{FKPP5} - \cite{FKPP9}.  

FKPP model is a simple reaction-diffusion equation that reads
\begin{equation}
\dot{\phi} = f(\phi) + D\phi'',
\end{equation}
where $f(\phi)$ is a growth model - for instance, logistic growth $f(\phi) = \delta \phi(1-\phi)$.  In the context of population dynamics, the density field $\phi$ represents the density (per unit length) of diffusing microorganisms - such as bacteria - which also reproduce themselves with rate $\delta$.   An isolated region of space inoculated at initial moment leads to the development of moving density fronts, known as Fisher waves.  

Recently, the role of delay in FKPP model has been examined \cite{DelayedFKPP1}-\cite{DelayedFKPP5}.  However, in these papers the delay has been introduced by modifying the growth function to have the form $\delta \phi(t)(1-\phi(t-\tau))$.   While the study of the dynamics of fronts that result from this is an interesting mathematical exercise, this is not necessarily the most natural way in which delay would arise physically.  
%

On the other hand, delay occurs naturally in scenarios that involves propagules and parent organisms.  In these situations, the particles do not reproduce themselves, and it would be unrealistic to apply FKPP model with a single density field to such biological settings.  Instead, propagules produce parent organisms, which, after a latent period produce new propagules, so this requires two separate density fields - for propagules and parent respectively.  Thus, a more appropriate basic model for this scenario would be as follows:
\begin{eqnarray}
\dot{\phi} &=& \delta \theta(t-\tau) + D\phi''  - \gamma \phi, \\
\dot{\theta} &=& \gamma \phi\left(1-\frac{\theta}{\theta_{max}}\right).
\end{eqnarray}
Here $\phi$ is the usual density of the propagules, while $\theta$ represents the density of a parent organism, such as fungi.  The parent is immobile - it grows on some substrate, such as soil.  However, the propagules can diffuse - again, reflecting the typical biological scenario in which this dynamics arises.  The rate $\gamma$ is the probability per unit time that a propagule turns into a parent - for example, a spore giving rise to a new fungus, and $\delta$ is the rate at which the parent substance produces propagules - for example, the rate at which fungi spew out spores.   Thus, the mobile propagules produce the immobile parent, which in turn produces new propagules.  

However, propagule production usually happens with a delay, because the parent needs to reach a certain stage of maturity before producing propagules.  The quantity $\tau$ is the delay time, commonly referred to in biological literature as ``latent time''.  A time $\tau$ must pass between the moment when a propagule has turned into a new parent substance and the moment when this parent begins to produce new spores.  In the mean time, new propagules are deposited.  Therefore, the rate of fungal production at time $t$ is not $\delta \theta(x,t)$, but $\delta \theta(x,t-\tau)$.  Finally, the parent substance typically has a carrying capacity - a maximal density $\theta_{max}$ at which it will stop growing.
%

In this paper we will investigate the behavior of density waves in this model, focusing in particular on the dependence of the speed of density fronts upon the model parameters.

First, we rescale variables to lower the number of relevant parameters.  Note that $\delta$ sets a natural timescale for this problem, namely $1/\delta$.  Therefore, we define dimensionless time by $\tilde{t} = \delta t$.  This is equivalent to measuring time not in seconds, but in units of $1/\delta$.  Also, the diffusion coefficient has dimensions of $\frac{distance^2}{time}$.  Therefore, a natural distance scale is $\sqrt{\frac{D}{\delta}}$, and we define a dimensionless coordinate $\tilde{x} = x\sqrt{\frac{\delta}{D}}$.  Finally, we define dimensionless densities by $\tilde{\phi} = \frac{\phi}{\theta_{max}}$ and $\tilde{\theta} = \frac{\theta}{\theta_{max}}$.  This is equivalent to expressing densities as a fraction of the maximal parent density.  Substituting all these variables we have 
\begin{eqnarray}
\label{eq:phi} \dot{\tilde{\phi}} &=& \tilde{\theta}(\tilde{t}-\mathcal{T}) + \tilde{\phi}'' - \Gamma \tilde{\phi}, \\
\label{eq:theta} \dot{\tilde{\theta}} &=& \Gamma \tilde{\phi}(1-\tilde{\theta}).
\end{eqnarray}
The derivatives are with respect to new, dimensionless variables.  We are left with only two parameters.  The first is the dimensionless infection rate $\Gamma$, which is given by $\Gamma = \gamma/\delta$ in terms of the physical parameters.  The second is the dimensionless latent time $\mathcal{T}$,  which is given by $\mathcal{T} = \tau\delta$ in terms of the physical parameters.  

The resulting speeds of fronts would also be dimensionless, $\tilde{v}$.  It will be a function of $\Gamma$ and $\mathcal{T}$.  To switch back to the original, physical variables, the dimensionless speed must be multiplied by $\frac{\sqrt{D/\delta}}{1/\delta} = \sqrt{\delta D}$.  In other words, the speed in physical units is given by 
\begin{equation}
\label{eq:dim_v}
v = \tilde{v}(\gamma/\delta, \tau\delta)\sqrt{\delta D}.
\end{equation}  
Therefore, we see that results scale like $\sim \sqrt{\delta D}$ - same as classical Fisher waves.  In the classical case, $\tilde{v}$ has a value $2$.  In the present model $\tilde{v}$ is a function of $\gamma/\delta$ and $\tau\delta$, and represents the departure from the classical FKPP result.  We examine this difference.

\section{Results}
\subsection{Front examples}
\begin{figure}[ht]
\includegraphics[width=3in]{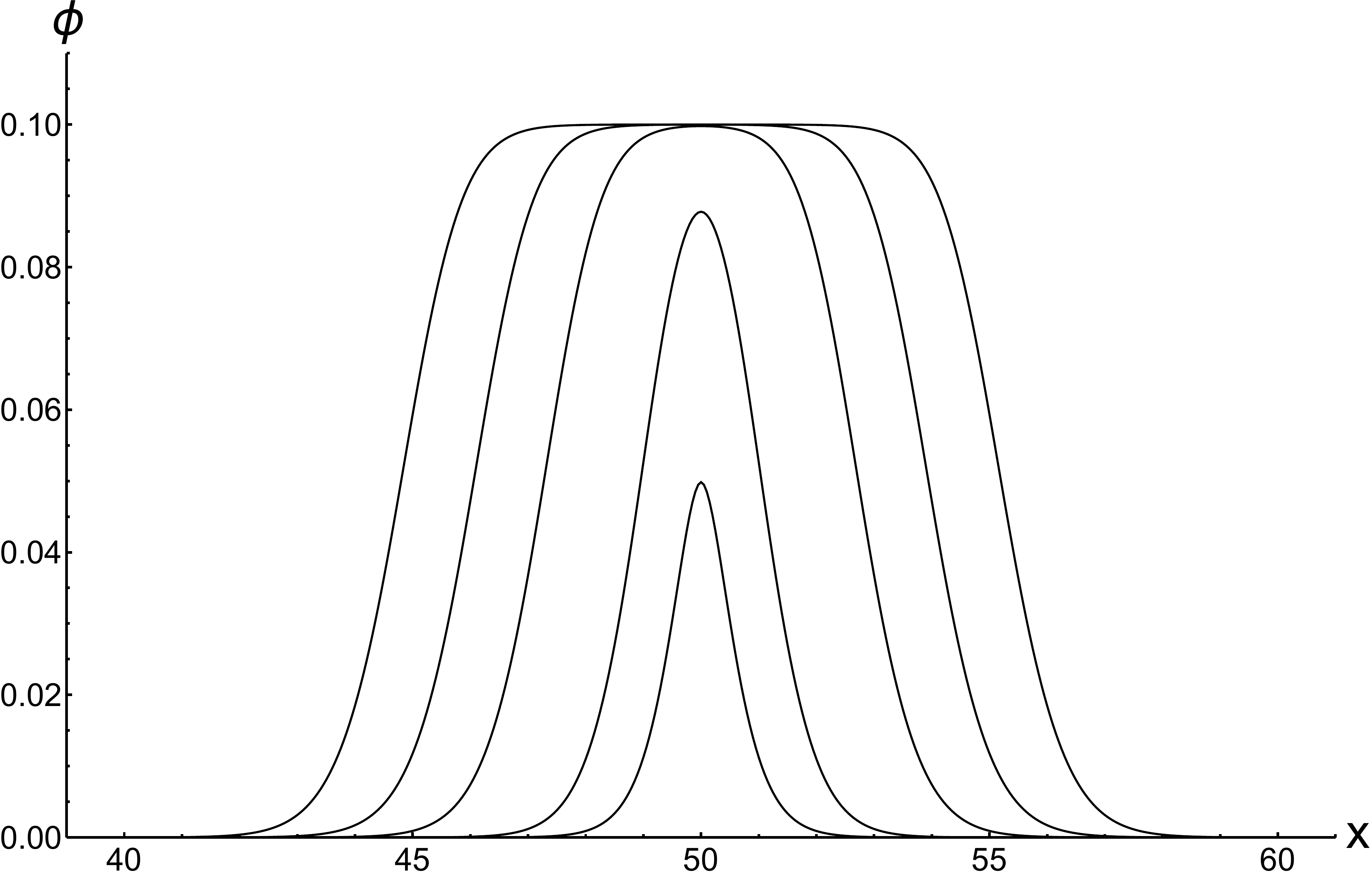}
\caption{Density wave $\phi$ vs. $x$ at $t=1.98$, $t=5.94$, $t=11.88$, $t=15.84$ and $t=19.8$.  The parameters are $\Gamma = 10$, $\mathcal{T}=1$.  The $\delta$-function initial condition at $t=0$ is not shown.}
\label{fig:phi_graph}
\end{figure}

\begin{figure}[ht]
\includegraphics[width=3in]{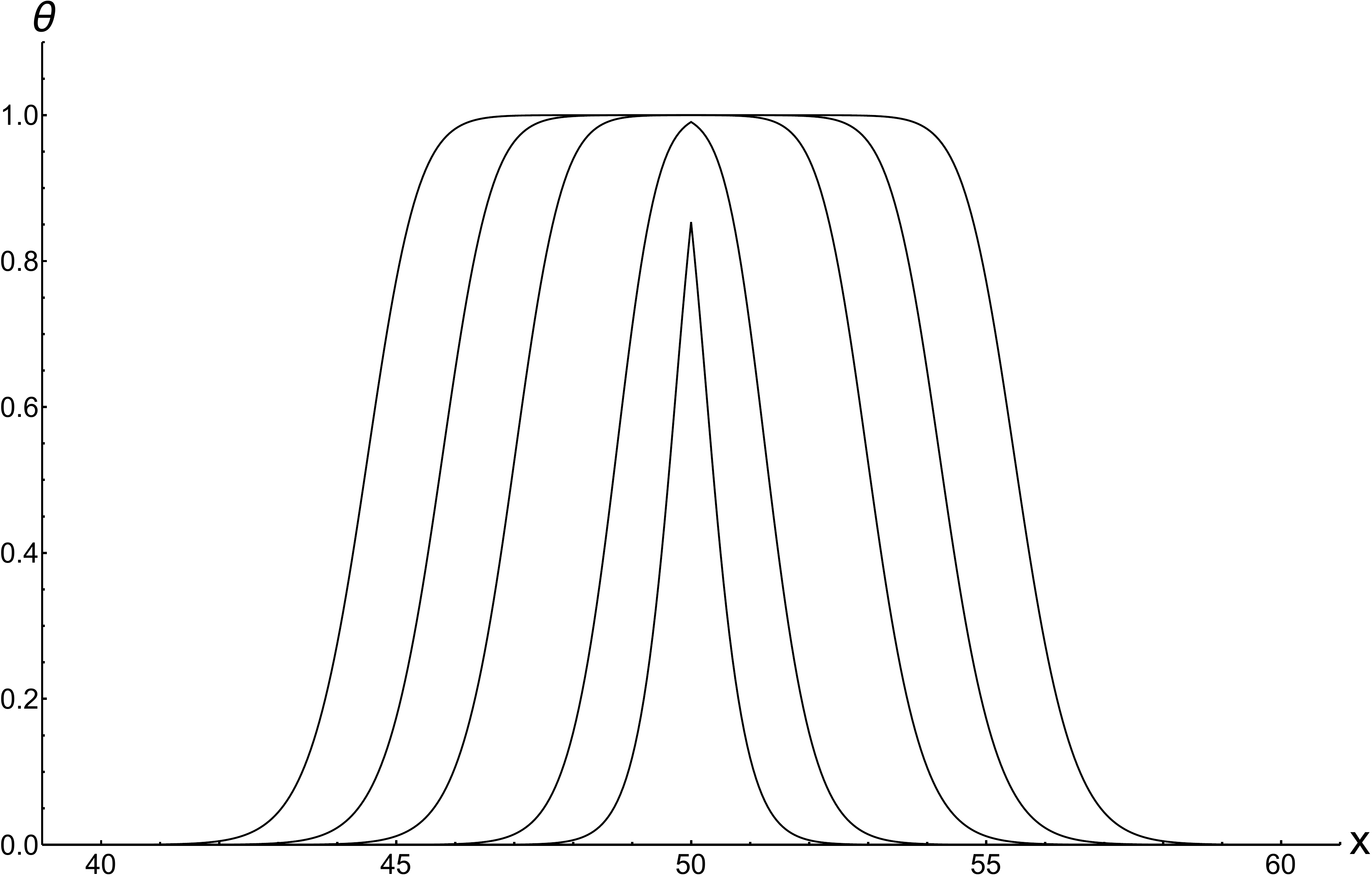}
\caption{Density wave $\theta$ vs. at $t=1.98$, $t=5.94$, $t=11.88$, $t=15.84$ and $t=19.8$.  The parameters are $\Gamma = 10$, $\mathcal{T}=1$.}
\label{fig:theta_graph}
\end{figure}

Figs.~\ref{fig:phi_graph} and \ref{fig:theta_graph} show examples of propagating density waves that result from a the initial condition $\phi(x,t=0) = \delta(x-x_0)$, and $\theta(x,t=0)=0$.  We see that after an initial transient, a pair of uniformly moving fronts are launched.  In the rest of the paper we will be concerned only with these uniformly moving fronts.
\subsection{Front speeds}
We now derive theoretical predictions for front speeds and compare them with numerical results.  This is a standard analysis that follows \cite{WvS}.  We will drop $\sim$ to lighten the notation, but will switch back as necessary when presenting results.  

First, we assume that a uniformly traveling front (UTF) exists, for which $\theta(x,t) = \theta(x-vt)$, and $\phi(x,t) = \phi(x-vt)$.  Thus, we seek a UTF solution by substituting this ansatz into equations of motion (\ref{eq:phi})-(\ref{eq:theta}).  We get
\begin{eqnarray}
-v\frac{d\phi}{dz} &=& \theta(z+v\mathcal{T}) + \frac{d^2 \phi}{dz^2} - \Gamma \phi, \\
-v\frac{d\theta}{dz} &=& \Gamma \phi(1-\theta),
\end{eqnarray}
where $z = x-vt$.  The UTF is described by a system of ordinary differential equations.  The equations have a fixed point at $(\theta=0, \phi=0)$ -- corresponding to the leading edge of the front, and a fixed point at $(\theta=1, \phi = 1/\Gamma)$ -- corresponding to a trailing edge of the front.  The solution near the $(0,0)$ fixed point determines the shape of the leading edge.

Next, we assume that fronts are pulled \cite{WvS}, which means that their speed is determined by the leading edge, where the nonlinear terms are negligible.  The pulled assumption will be validated through comparison with numerical predictions of Eqs.~(\ref{eq:phi})-(\ref{eq:theta}).  Thus, the leading edge is described by the linearized equations  
\begin{eqnarray}
-v\frac{d\phi}{dz} &=& \theta(z+v\mathcal{T}) + \frac{d^2 \phi}{dz^2} - \Gamma \phi, \\
-v\frac{d\theta}{dz} &=& \Gamma \phi.
\end{eqnarray}
The solution will have the form $\phi = \phi_0 e^{-\lambda z}$, and $\theta = \theta_0 e^{-\lambda z}$.  Substituting this into the linearized equations produces a relationship
\begin{equation}
\label{eq:slambda_eq}
\Gamma e^{-v\lambda \mathcal{T}} - (v\lambda)^2 - v\lambda \left(\Gamma-\lambda^2 \right) =0,
\end{equation}
which relates the front speed $v$ and the decay length of the front $\lambda$, i.e. it gives a function $v(\lambda)$.  The theory of \cite{WvS} states that for initial conditions (IC) that decay in space faster than an exponential with a certain critical decay rate, the UTF solution will have a $\lambda$ that minimizes $v(\lambda)$.  All localized ICs, such as a $\delta$-function IC satisfy this criterion.  This minimum will take place at $\lambda^*$, and the front speed will be $v^* = v(\lambda^*)$.  

\begin{figure}[ht]
\center \includegraphics[width=3.4in]{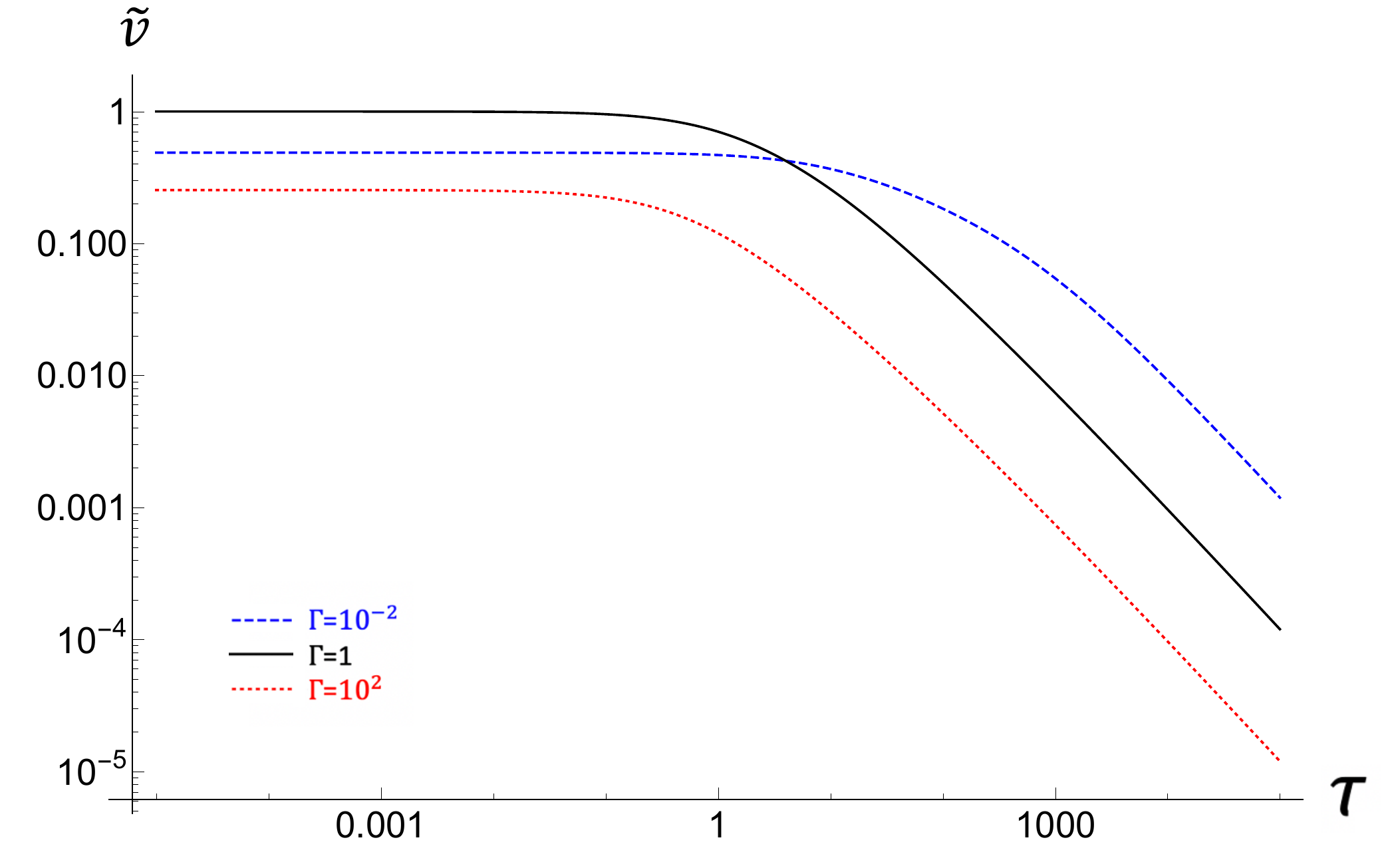}
\caption{(Color online) Dimensionless speed $\tilde{v}$ vs.~dimensionless latent times $\mathcal{T}$ for various dimensionless infection rates $\Gamma$.}
\label{fig:Result}
\end{figure}

\begin{figure}[ht]
\center \includegraphics[width=3.4in]{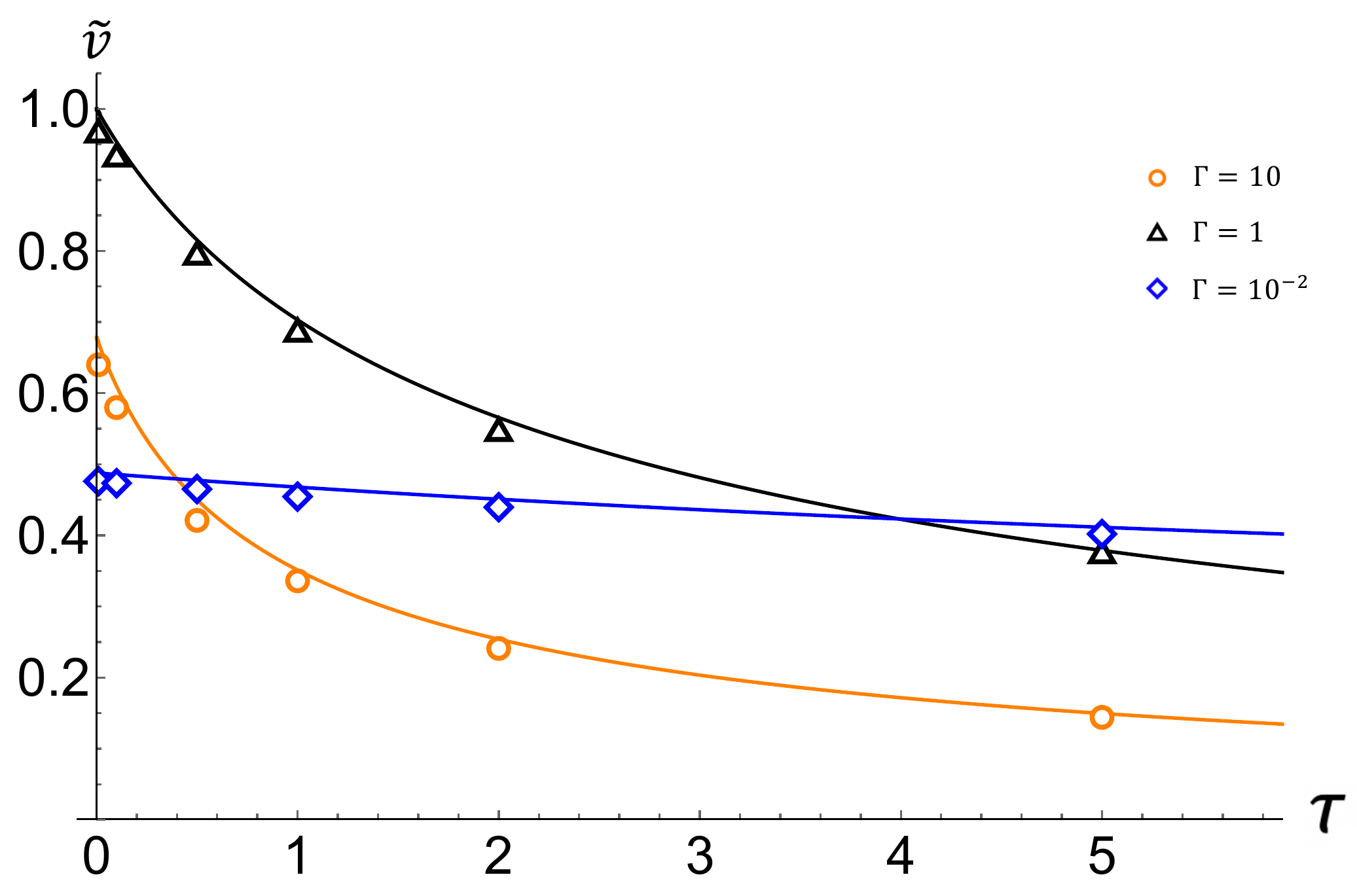}
\caption{(Color online) Comparison of theoretical $\tilde{v}(\mathcal{T})$ with values extracted from the numerical solution of Eqs.~(\ref{eq:phi})-(\ref{eq:theta}) for three different values of $\Gamma$.}
\label{fig:TheoryVsNumerics}
\end{figure}

Taking the derivative of Eq.~(\ref{eq:slambda_eq}) and solving for $v'(\lambda)$ gives
\begin{displaymath}
v'(\lambda) = -\frac{\Gamma + \lambda v(\lambda)(\Gamma \mathcal{T} - \lambda e^{\lambda \mathcal{T} v(\lambda)}(2\lambda-v(\lambda)))}{\lambda(\Gamma + \Gamma \mathcal{T} \lambda v(\lambda) + \lambda^2 s^2(\lambda) e^{\lambda \mathcal{T} v(\lambda)} }
\end{displaymath}
This is zero when the numerator is zero.  Thus, we can in principle solve for $v(\lambda)$ from Eq.~(\ref{eq:slambda_eq}), substitute into the numerator, set it to zero, and find $\lambda^*$.  Alternatively, we can treat $v^*$ and $\lambda^*$ as two independent variables and find them from a simultaneous solution of Eq.~(\ref{eq:slambda_eq}) and the numerator$=0$, i.e. 
\begin{eqnarray}
\label{eq:star1} &&\Gamma e^{-v^*\lambda^* \mathcal{T}} - (v^*\lambda^*)^2 - v^*\lambda^* \left(\Gamma-(\lambda^*)^2 \right)=0,\\
\label{eq:star2} && \Gamma + \lambda^*v^*(\Gamma \mathcal{T} - \lambda^* e^{\lambda^* \mathcal{T} v^*}(2\lambda^*-v^*)) = 0.
\end{eqnarray}
This was done numerically for a set of $\Gamma$ and $\mathcal{T}$.  Reverting back to the dimensionless notation, this produces $\tilde{v}(\Gamma, \mathcal{T})$.  The result is plotted in Fig.~\ref{fig:Result} for several $\Gamma$ and $\mathcal{T}$.  Fig.~\ref{fig:TheoryVsNumerics} compares these theoretical predictions with front speeds obtained from the numerical solution of Eqs.~(\ref{eq:phi})-(\ref{eq:theta}).  The details of our numerical approach can be found in the Appendix.
\subsection{Analysis}
Fig.~\ref{fig:Result} demonstrates that our results have three main features.  The first, is the limiting speed as $\mathcal{T}$ goes to zero.  
The second, is the large $\mathcal{T}$ asymptotic regime, where $v(\mathcal{T})$ appears to have an asymptotic behavior reminiscent of a power law (we show below that it is not a pure power law).  The the third, is the characteristic crossover point beyond which this asymptotic approximation is valid.  This crossover point is a function of $\Gamma$.  We now extract these properties.

\subsubsection{Zero delay}
\begin{figure}[ht]
\center \includegraphics[width=3.4in]{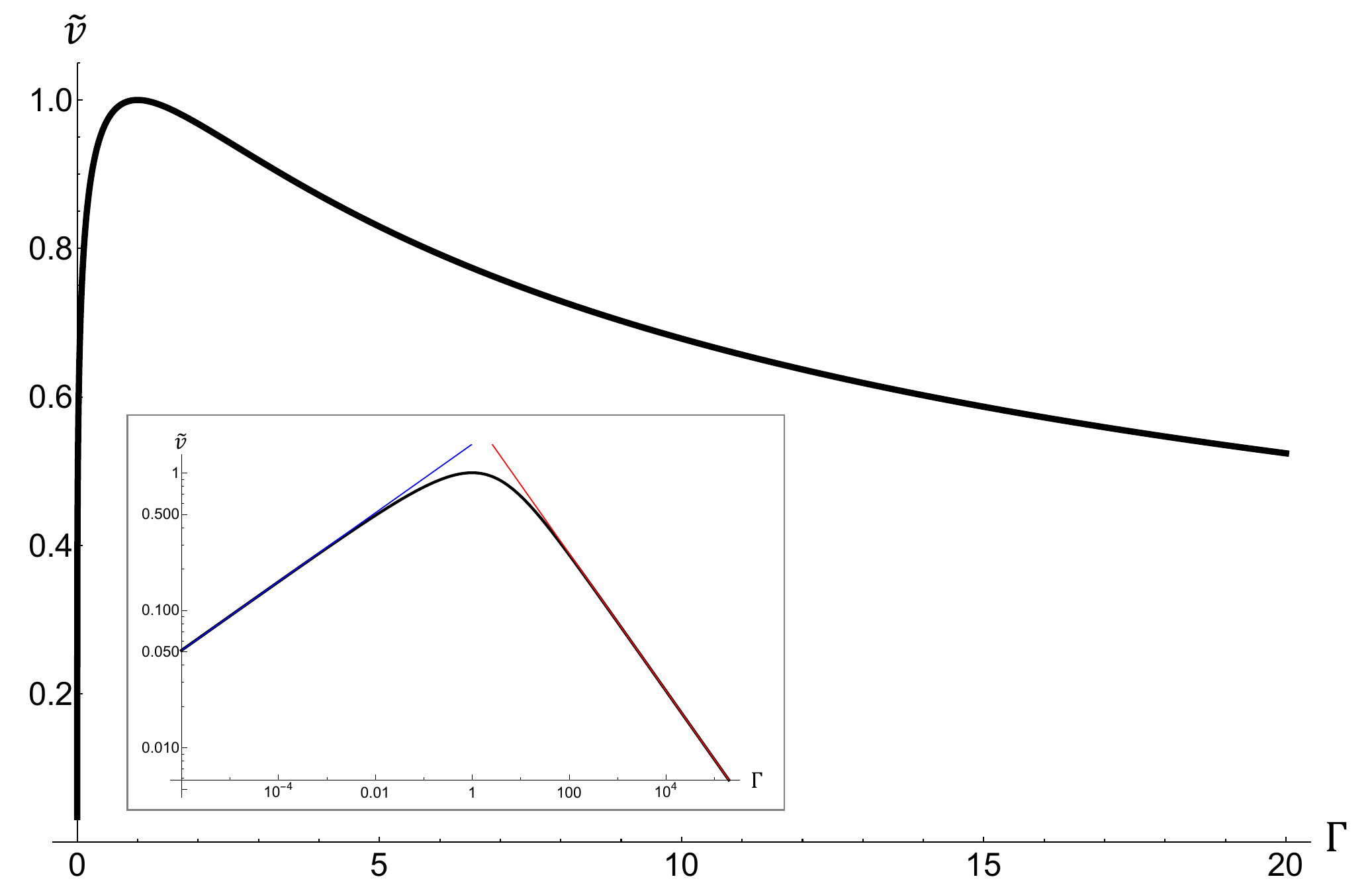}
\caption{(Color online) Dimensionless speed $\tilde{v}(\Gamma)$ at zero delay $\mathcal{T}$.  Inset: The same quantity on a log-log scale.  The two thin lines are the low- and high- $\Gamma$ asymptotics, $\frac{3^{3/4}}{\sqrt{2}}\Gamma^{1/4}$ (blue) and $2.598 \Gamma^{-1/2}$ (red) respectively.}
\label{fig:ZeroDelay}
\end{figure}

First, we study the speed at zero delay.  Setting $\mathcal{T}=0$ in Eqs.~(\ref{eq:star1})-(\ref{eq:star2}) we can produce an analytic result.  There are four roots, but only one of which is positive and real.  Reverting back to the dimensionless notation, this root is given by 
\begin{equation}
\label{eq:ZeroTau} \tilde{v}(\Gamma) = \frac{\left(6+\Gamma - \sqrt{\Gamma(3+\Gamma)}\right)\sqrt{-3\Gamma + 6\sqrt{\Gamma(3+\Gamma)}}}{3(4+\Gamma)}.
\end{equation}
This function is plotted in Fig.~\ref{fig:ZeroDelay}.  
At small $\Gamma$ this function behaves as $\frac{3^{3/4}}{\sqrt{2}}\Gamma^{1/4}$.  At large $\Gamma$, it approaches the asymptotic behavior $\frac{3^{3/2}}{2}\Gamma^{-1/2}$.  The speed reaches the maximum value $1$ at $\Gamma=1$.  

At $\Gamma=0$, there is no spore production, so the model reduces to a pure diffusion equation with no growth, lacking fronts;  this is the meaning of zero front speed.  On the other hand, as $\Gamma$ approaches infinity we also do not recover FKPP model, because while the rate of conversion from propagules to the substrate density goes to infinity, the rate of production of new propagules remains $1$.  Therefore, this is effectively equivalent to killing of propagules with infinite rate, giving a speed zero.  So, we see that the model does not match to KFPP model in either limit.  The optimal speed is achieved at intermediate $\Gamma=1$, but theres no reason for it to reach the KFPP value of $\tilde{v}=2$, since in this intermediate regime the model is clearly not equivalent to the FKPP model.   

Evidently, this model does not reproduce FKPP results for any parameters, and there is no reason to expect a correspondence, because the propagule production terms are different.  For instance, the localized dynamics in the absence of transport is a first-order in time in the classical FKPP case, but it is second-order in time in the new model.  However, as we have shown, both models predict $v \propto \sqrt{\delta D}$, so it is meaningful to compare the proportionality factor, which is $2$ in the classical model, but depends on the parameter $\gamma$ in the new model.

\subsubsection{Finite delay}
Turning on a finite delay slows down the front, as we can see from Fig.~\ref{fig:Result}.  To understand the large-$\mathcal{T}$ asymptotic behavior, we noticed from numerical calculations of solutions to Eqs.~(\ref{eq:star1}) -(\ref{eq:star2}) that the product $v^* \lambda^* \mathcal{T}$ tends to a number greater than $1$ as $\mathcal{T} \rightarrow \infty$ for any $\Gamma$.  Also, $v^*$ goes to zero, while $\lambda^*$ does not.  With the help of this observation, Eqs.~(\ref{eq:star1})-(\ref{eq:star2}) at large $\mathcal{T}$ read
\begin{eqnarray}
\label{eq:star1_asymptotic} &&(v^*\lambda^*) - \Gamma+(\lambda^*)^2 = 0,\\
\label{eq:star2_asymptotic} && \Gamma - 2v^*(\lambda^*)^3 e^{\lambda^* \mathcal{T} v^*} =0.
\end{eqnarray}
\begin{figure}[ht]
\center \includegraphics[width=3.4in]{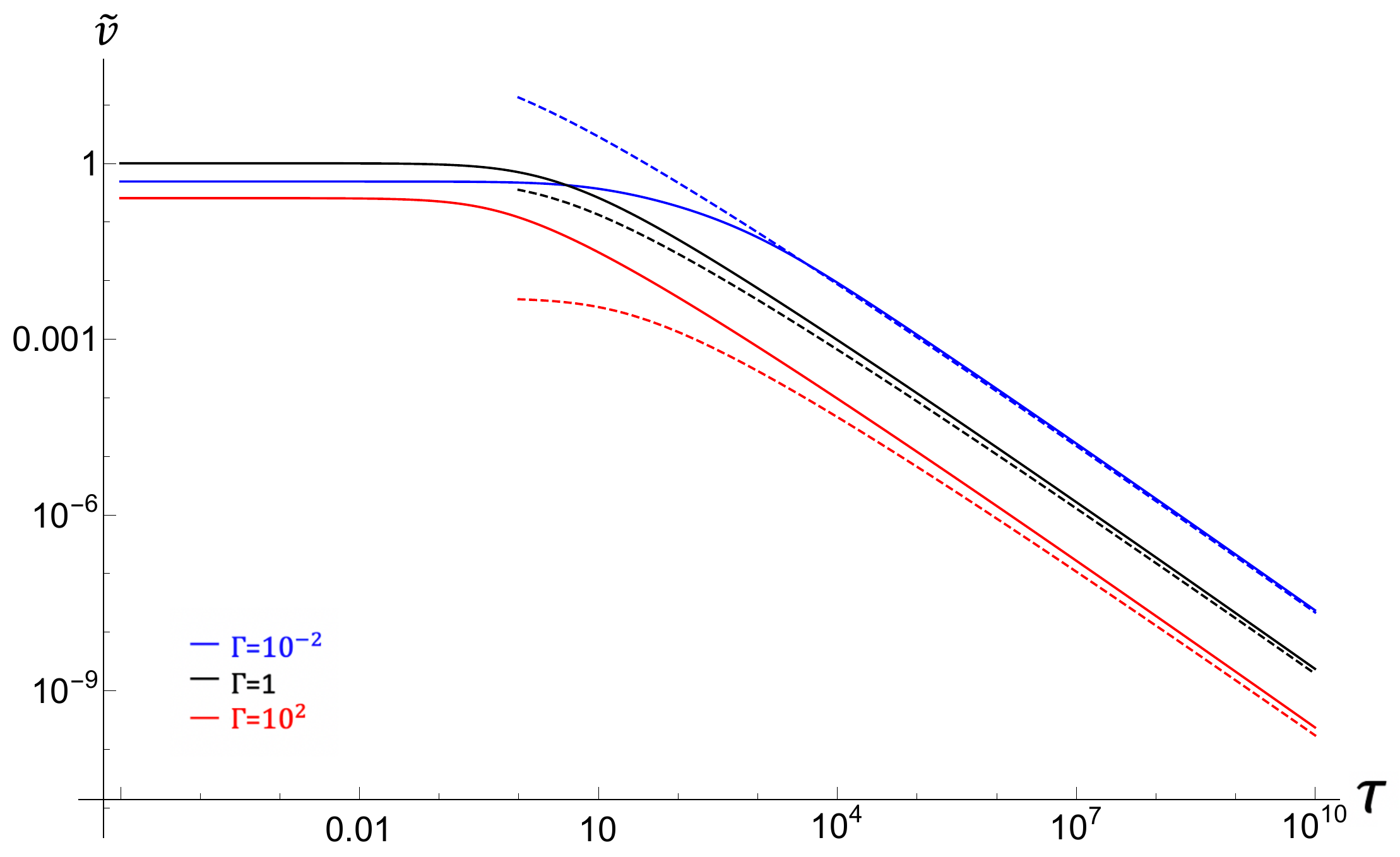}
\caption{(Color online) Tails of $\tilde{v}(\mathcal{T})$ given by Eq.~(\ref{eq:tail_equation}) -- dashed curves, compared with the exact result, obtained from Eqs.~(\ref{eq:star1})-(\ref{eq:star2}) -- solid curves.}
\label{fig:Tails}
\end{figure}
The first equation readily gives $\lambda^* = \frac{1}{2}\left(v^*\pm \sqrt{(v^*)^2+4\Gamma}\right)$, and we must chose the positive root, because only this choice always produces a positive $\lambda^*$.  It is convenient to factor out the quantity $2\sqrt{\Gamma}$ from the speed, so we let $v^* = 2\sqrt{\Gamma} w$.  Substituting both of these expressions into the second equation gives
\begin{equation}
4\Gamma^{3/2} w\left(w+\sqrt{w^2+1}\right)e^{(2\Gamma \mathcal{T}) w\left(w+\sqrt{w^2+1}\right)} =1
\end{equation}
Now, an equation of the form $1=aue^{bu}$ has a solution $u=\frac{W(b/a)}{b}$, where $W$ is Lambert \emph{W} function, or product logarithm.  Therefore, we have
\begin{equation}
w\left(w+\sqrt{w^2+1}\right) = \frac{W\left(\frac{\mathcal{T}}{2\sqrt{\Gamma}}\right)}{2\Gamma\mathcal{T}}.
\end{equation}
This has a solution
\begin{displaymath}
w =\frac{W\left(\frac{\mathcal{T}}{2\sqrt{\Gamma}}\right)}{2\Gamma \mathcal{T}\sqrt{1+\frac{W\left(\frac{\mathcal{T}}{2\sqrt{\Gamma}}\right)}{\Gamma\mathcal{T}}}} \approx \frac{W\left(\frac{\mathcal{T}}{2\sqrt{\Gamma}}\right)}{2\Gamma \mathcal{T}}.
\end{displaymath}
Thus, the high-$\Gamma$ tail of the dimensionless speed, in the dimensionless notation, is 
\begin{equation}
\label{eq:tail_equation}
\tilde{v} = \frac{W\left(\frac{\mathcal{T}}{2\sqrt{\Gamma}}\right)}{\sqrt{\Gamma} \mathcal{T}}
\end{equation}
We compare predictions of this equation with exact results in Fig.~\ref{fig:Tails}.  Evidently, the tail behavior of $\tilde{v}(\mathcal{T})$ is not a pure power law.  We can estimate the crossover point approximately by equating the value at zero $\tau$ given by Eq.~(\ref{eq:ZeroTau}) and the denominator of Eq.~(\ref{eq:tail_equation}).  The result is a complicated function that diverges at small $\Gamma$ as $\frac{\sqrt{2}}{3^{3/4}} \Gamma^{-3/4}$ and tends to a value $\approx 0.3849$ as $\Gamma \rightarrow \infty$.

\section{Conclusion}
The largest speed in our model is $\sqrt{\delta D}$, which is achieved at zero delay and when the rate of conversion from propagules to parent is matched to the rate at which the parent produces propagules.  This speed is exactly $1/2$ of the classical Fisher speed, although, one must be careful to take this comparison at face value, since the terms that model production of new material are different in the two models.  While propagules are produced by a parent organism in the new model, they are produced by other propagules in the basic FKPP model.  Still, the speed of both models is proportional to $\sqrt{\delta D}$, but they differ in proportionality constants.  

Thus, we find that invasion fronts in organisms that require a parent for reproduction will move at most half as slowly as invasion fronts in organisms that reproduce themselves.  All other common rates being equal, invasion bacterial fronts will move twice as fast as invasion fungal fronts.  

When the rate of propagule production by the parent does not equal to the rate of conversion from propagules to the parent, the front speed is slower than this optimal value.  Increasing conversion rate past the production rate decreases front speed, and for sufficiently large conversion rate, the speed scales as conversion rate to the power $-1/2$.  
This is reminiscent of an effect was recently described in a completely different field.  The authors of \cite{QuantumZeno} describe an open quantum system of free fermions driven by a source that injects fermions.  When the injection rate is sufficiently large, the rate at which the number of fermions in the system grows with time scales inversely with the injection rate.  The authors explain that this is a manifestation of Quantum Zeno effect.  Moreover, our Fig.~\ref{fig:ZeroDelay} over the entire range of $\Gamma$ resembles Fig.~1 in \cite{QuantumZeno} over the entire range of injection rates.  

As one would expect, the delay further slows down invasion fronts.  We find that for sufficiently large delay $\tau$, the front speed scales approximately like $1/\tau$, although a more exact result multiplies this by a product logarithm function with an argument linear in $\tau$.  

\section*{Acknowledgements}
We would like to thank William and Linda Frost Fund fund for supporting undergraduate research at Cal Poly.  

\appendix
\section{Numerical Method}
\label{sec:Numerical}
Our numerical method uses a first-order time-differencing scheme, for example
\begin{equation}
\left. \frac{\partial \phi}{\partial t}\right|_{x_,t_n}  \rightarrow \frac{\phi(x_m,t_{n+1}) - \phi(x_m,t_n)}{\Delta t}
\end{equation}
and an upwind spatial differencing scheme,
\begin{equation}
\left.\frac{\partial \phi}{\partial x}\right|_{x_m,t_n} \rightarrow \frac{\phi(x_m,t_n) - \phi(x_{m-1},t_n)}{\Delta x},
\end{equation}
applied to Eqs.~(\ref{eq:phi})-(\ref{eq:theta}).  Here $x_m = x_0 + m\Delta x$ and $t_n = t_0 + n\Delta t$.  
\begin{figure}[ht]
\includegraphics[width=3in]{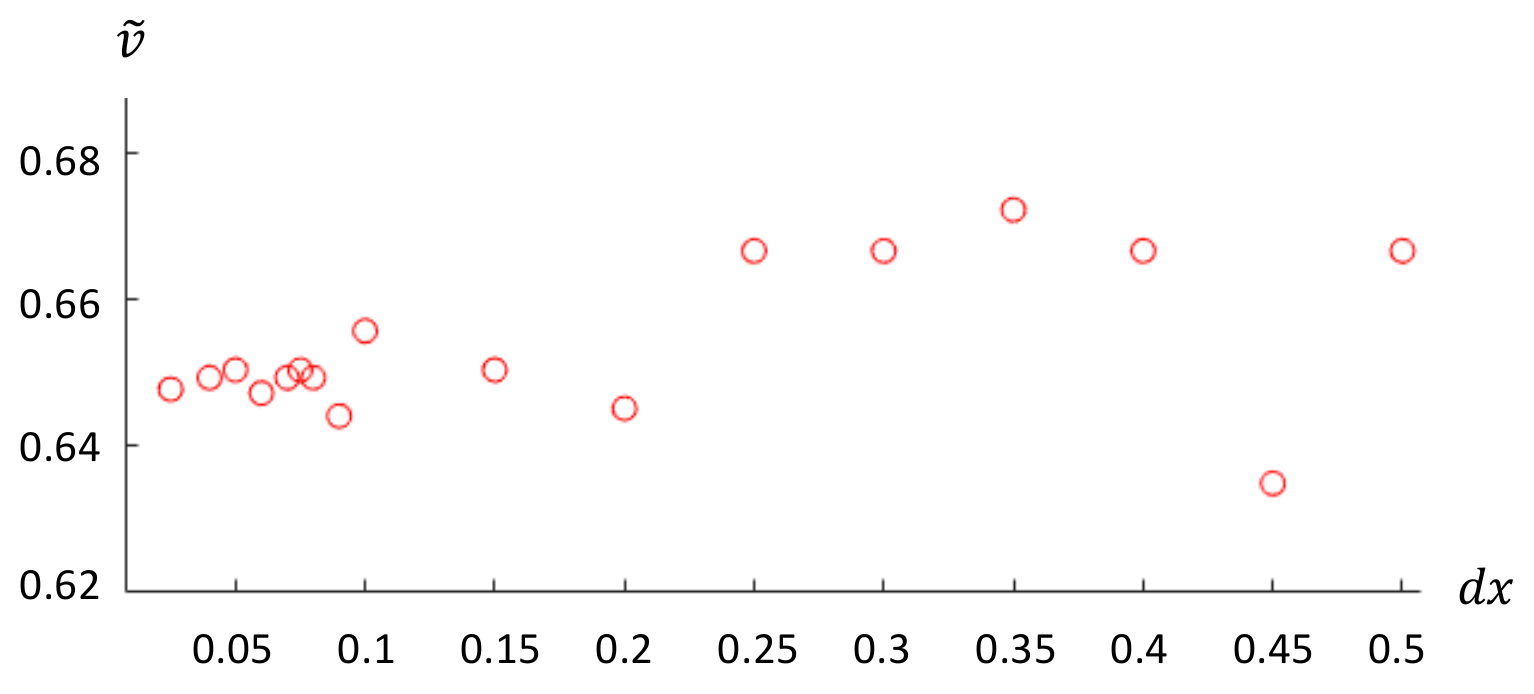}
\caption{Front speed as a function of $dx$ for $\tau = 1$, $\Gamma=1$, and simulation time $T=15$.}
\label{fig:dx_dependence}
\end{figure}
The integration was performed over a finite spatial interval, with an initial condition placed in the center.  At the extreme left and extreme right points of this spatial interval, we set the values of both densities to zero.  However, we the spatial domain was chosen large enough so that neither front ever came close to either boundary given the placement of the initial profile.

We performed convergence test for select parameter combinations.  An example is shown in Fig.~\ref{fig:dx_dependence}.   The front speed generated by the simulation code does not show a strong dependence on $\Delta x$.  All the results presented in the paper were obtained with $\Delta x=0.05$, which is much less than the typical characteristic width of fronts.  To ensure stability, we chose $\Delta t = 0.25 (\Delta x)^2$ as suggested by an earlier paper on a similar model \cite{CB}.  Instabilities were not observed.  To extract the speed of the wavefronts from these simulations, we tracked the position of the contour of a fixed reference density, chosen to lie in the leading edge.  There is a transient time period during which the shapes of fronts develop.  During this same transient period the speed of fronts also changes.  As time progresses, the front profiles tend to a stationary shape, and front speeds tend to a limiting value.  This limiting value will be identical for both densities, $\phi$ and $\theta$, although instantaneous front speeds of the two densities is generally not identical during the transient period.  

\begin{figure}[ht]
\center \includegraphics[width=3in]{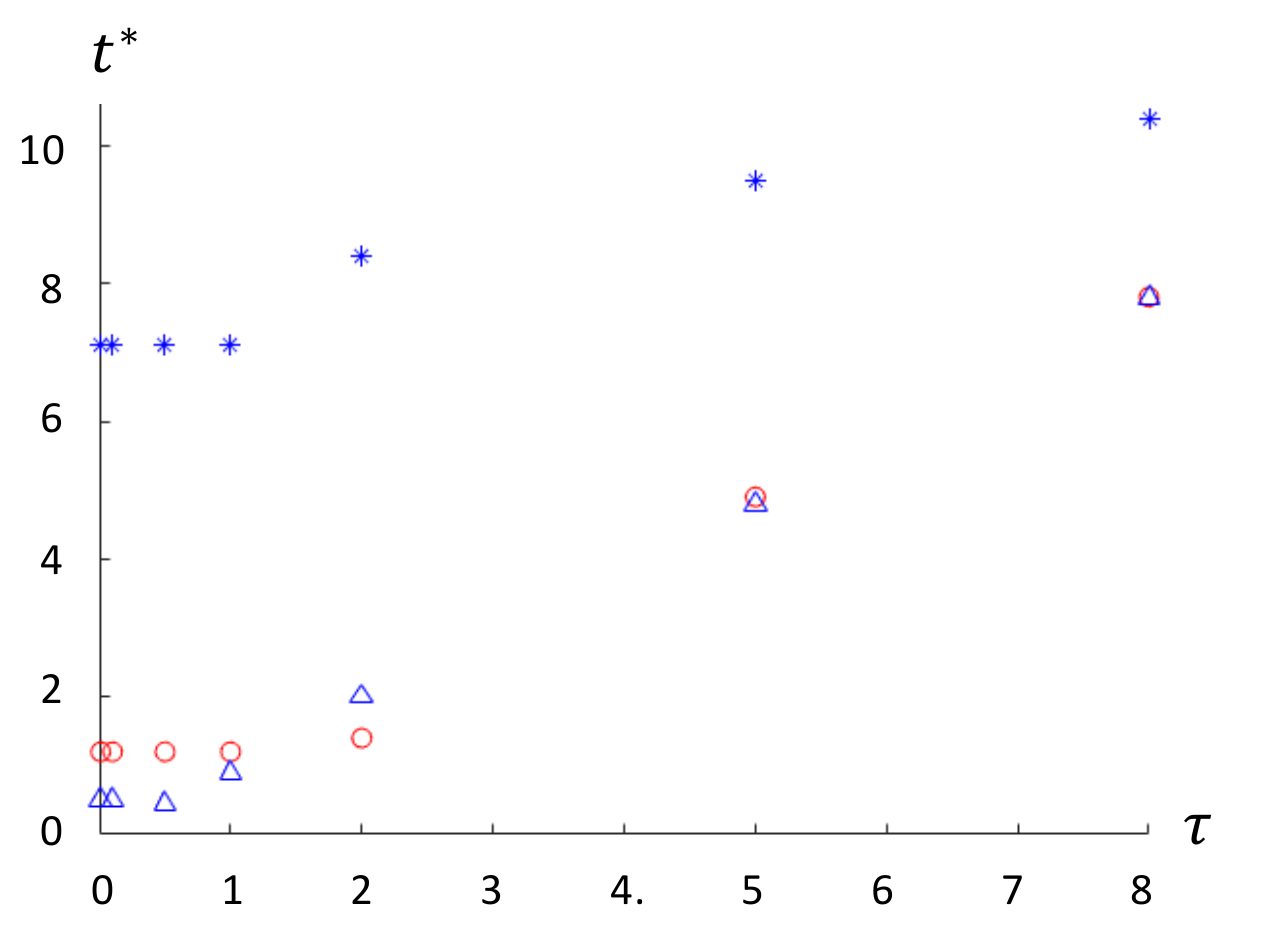}
\caption{(Color online) $t^*$ as a function of $\tau$.  Blue asterisks: $\Gamma=10^{-2}$, red circles: $\Gamma=1$, blue triangles: $\Gamma=10$.}
\label{fig:tstarvstau}
\end{figure}

\begin{figure}[ht]
\center \includegraphics[width=3in]{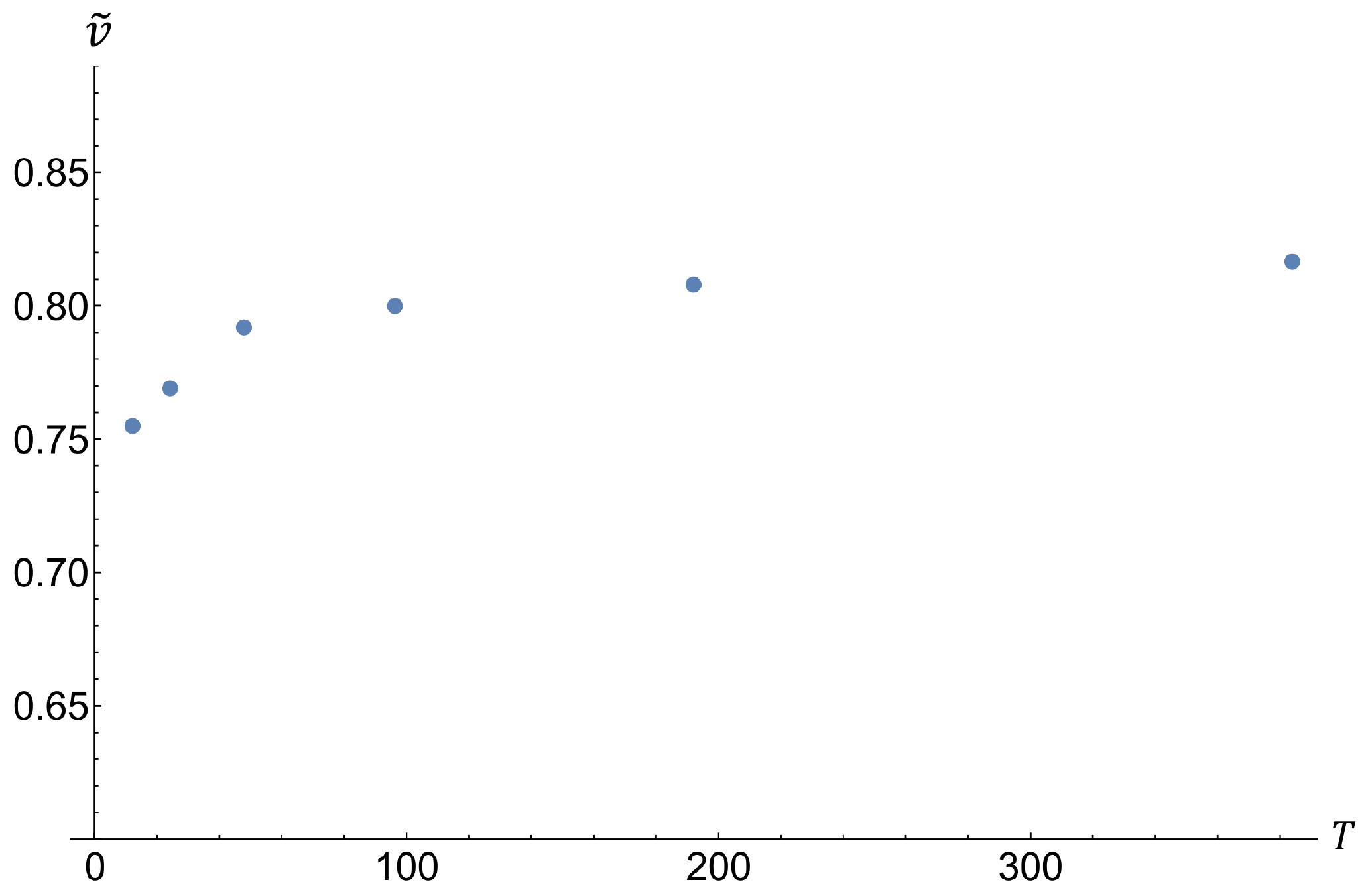}   
\caption{Front speed $\tilde{v}$ versus the total simulation time $T$ for $\Gamma=1$, and $\tau=0.5$.  The data points are calculated for $T=10t^*$, $T=20t^*$, $T=40t^*$, $T=80t^*$, $T=160t^*$, $T=320t^*$.}     
\label{fig:saturation}
\end{figure}

The rate of convergence to the limiting speed depends on parameters.  We needed to develop a consistent method of scaling the total run time $T$ with this convergence rate, and to ensure the simulations ran long enough for density profiles to come close enough to the asymptotic speed.  In order to make this choice, we used the following idea.  Starting from a $\delta$-function IC, the $\phi$ profile very quickly diffuses and the maximal value rapidly decreases.  At a certain instant of time $t^*$ it reaches the smallest value, after which the $\phi$ profile grows again.  This is a characteristic time at which the growth mechanism becomes important.  We display $t^*$ vs $\tau$ in Fig.~\ref{fig:tstarvstau}.
We used $t^*$ as a characteristic measure of the transient time.  Fig.~\ref{fig:saturation} demonstrates the measured speed of the front as a function of the total simulation time $T$, expressed as multiples of $t^*$ for a particular $\Gamma$ and $\tau$.
Here, we see that the front speed maintains an upward trend for greater $T$ and saturates to an equilibrium value for sufficiently large $T$.   We then scaled the simulation time to be $T=40t^*$.  The number $40$ was chosen as a compromise between a sufficiently small error from the theory, and a practical simulation time, since increasing $T$ increases the simulation time dramatically.  We repeated the procedure for $\Gamma = 10^{-2}$ and $\Gamma=10$,  although $20t^*$ was used for $\Gamma=10^{-2}$ - it gave a sufficient accuracy in comparison with the theoretical curve.  The linear trend in the function $t^*(\tau)$ can be interpolated to choose the appropriate $T$ at any $\tau$ in this linear region.


\end{document}